\documentclass[twocolumn,showpacs,amsmath,amssymb,prb]{revtex4}
\usepackage{epsfig}

\begin{document}
\title{Midgap States and Generalized Supersymmetry in Semi-infinite Nanowires}
\author{Bor-Luen Huang$^{1}$, Shin-Tza Wu$^{2}$,
and Chung-Yu Mou$^{1,3}$}
\affiliation{
1. Department of Physics, National Tsing Hua University, Hsinchu 30043,
Taiwan \\
2. Department of Physics, National Chung-Cheng University, Chiayi 621, Taiwan\\
3. National Center for Theoretical Sciences, P.O.Box 2-131, Hsinchu, Taiwan}
\date{\today}

\begin{abstract}
Edge states of semi-infinite nanowires in tight
binding limit are examined. We argue that understanding
these edge states provides a pathway to generic comprehension
of surface states in many semi-infinite physical systems.
It is shown that the edge states occur within the gaps of the
corresponding bulk spectrum (thus also called the midgap states).
More importantly, we show that the
presence of these midgap states
reflects an underlying generalized supersymmetry.
This supersymmetric structure is a generalized
rotational symmetry among sublattices and results in a 
universal tendency: all midgap states tend to
vanish with periods commensurate with the underlying lattice.
Based on our formulation, we propose a structure with superlattice in hopping
to control the number of localized electronic states occurring at the ends
of the nanowires.
Other implications are also discussed.
In particular, it is shown that
the ordinarily recognized
impurity states can be viewed as disguised midgap states.
\end{abstract}


\maketitle

\section{Introduction}
\label{Int}
The one-dimensional (1D) wire has been of great
theoretical and experimental interests in the past. This is because of not only
the wide variety of fascinating phenomena it exhibits, but also the
testing grounds it offers for ideas that may become applicable in
higher dimensions. 
In practice, 1D wires needs not be physically 
one dimension. It may result from
projection after a partial Fourier transformation
from higher dimensional models.
For instance, a superlattice structure can be
reduced to an equivalent 1D structure
after a partial Fourier transformation along
the direction normal to the layers.
Similar examples include $d$-wave superconductors,
graphite sheet, and many other systems.
Therefore, understanding the 1D wire is an ideal
first step toward the understanding of
any higher dimensional problems. Further boost for
studying 1D wires comes from recent advances achieved 
in nanotechnology.
Here the feasibility for bottom-up assembly of single nanowires\cite{Ho}
has made direct investigation of
finite 1D wires possible. 
Nevertheless, conventional studies of the 1D wire have mostly been
focused on its bulk properties, whereas assembled nanowires can
only have finite lengths and must terminate at some sites (the ends,
or the edges). It is therefore desirable to reconsider the effects of
the ends to the properties of the nanowires.

The commonly recognized edge (or surface) effects in the physics of
nanostructures are concerned with the large volume fraction of the boundaries.
However, from either fundamental or practical viewpoints,
the possible occurrence of edge
modes and its influences on the properties of the system poses a
much more interesting problem.
For example, when applying carbon nanotubes
as emitters for screen displays, the occurrence of edge states may change the
density of electrons at the edge and thus affects the threshold
working potential. It is therefore of great technological interest if one
could devise a way to engineer the number of edge states.
From the fundamental viewpoint, the elegant role of edge excitations in
the physics of Quantum Hall systems\cite{wen} is a well known example
that illustrates the importance of edge states.
Generally speaking, the edge states occur within the gap of the
bulk energy spectrum and are called the midgap states. The existence of
these states causes anomalous properties near the end which can manifest
in tunneling measurements. A recently discovered
example is the
zero-bias conductance peak observed in the $dI/dV$ measurement of the
metal-$d$-wave superconductor junctions\cite{mou1}.
When electron-electron interactions are present,
as occurs for an externally-implemented magnetic impurity,
"intrinsic" Kondo effects may also arise due to these localized states,
causing zero-bias anomaly near the Fermi energies
\cite{kondo}. Furthermore, if the system is finite,
coupling between edge states can not be neglected.
An example is the anomalous paramagnetic behavior observed in
carbon nanoribbons\cite{ribbons}, where we have recently shown that
there are residual antiferromagnetic couplings
between edge spins in this system\cite{mou}.
All these examples clearly illustrate the important role of edge states in the
physics and applications of nanostructures.

In previous work, applying the Green's function approach,
we have shown that broken reflection symmetry
is a necessary condition for the occurrence of edge states,
and the energies of edge states are the roots to the Green's function
\cite{mou1}. In this
work, resorting to the supersymmetric method, 
we further develop a
systematic way to determine the wavefunctions
and the precise energies of the edge states.

Conventionally, the usage of the supersymmetric method 
in the condensed matter physics
has been focused on applying the supersymmetry (SUSY) quantum field theory 
to disorder
systems\cite{susy0}.
The application of the corresponding (0+1) dimensional limit
-the SUSY
quantum mechanics\cite{susy1}, however,
is quite limited. Nevertheless,
it has been realized\cite{Legget}
that the zero-bias
anomaly in $d$-wave superconductors
is closely related with the SUSY quantum mechanics.
These studies, however, are done in
in the continuum limit, using the semi-classical approximation,
while the more relevant limit
for high Tc superconductors and many
other systems is the tight binding limit.
Furthermore, the zero-energy state was the primary focus, while
not all the states localized at the edge have zero energy. 
It is therefore important to see if  
the idea of SUSY quantum mechanics can be generalized to
understand the finite-energy midgap states, in particular those
in the discrete condensed matter systems, as well.
In this work, we shall show that indeed, this is possible.
We shall first show that the semi-infinite 
tight-binding $d$-wave superconductors belongs to a 
more general class, the bipartite system, and which
can be well
described by the  
conventional SUSY quantum mechanics\cite{susy1}.
Here the supersymmetric partners
are two sublattices of the same system and
the SUSY is characterized by a hermitian supercharge $Q$
and the SUSY Hamiltonian $H^S=\{Q,Q\} /2$ with $[H^S,Q]=0$.
For bipartite systems with
nearest neighbor hoppings,
$Q$ is identical to the physical Hamiltonian ($\equiv H_2$) 
and hence $H^S$ is a quadratic functional of $H_2$.
Furthermore, the zero-energy state is annihilated by
the supercharge, which then constitutes one of the conditions 
for determining the zero-energy state; while the
other condition is to require it to decay from the edge. 
It is found that this conventional SUSY quantum mechanics
can be appropriately extended to describe the
semi-infinite $p$-partite 
systems with
nearest neighbor hoppings.
First, when $p \geq 3$, the original supercharge
splits into two: In addition to the physical
Hamiltonian $H_p$, a second supercharge $Q_p$
can be formed. 
They both commute with 
the SUSY Hamiltonian $H_S$. Furthermore, only
when $p=2$, $H_S \approx H^S$ is a quadratic functional of
$H_p$. In general,
$H_S$ is a polynomial functional
of $H_p$. This is a reminisce of the fractional
SUSY quantum mechanics\cite{fractional} in which
the SUSY Hamiltonian is generalized to be integer power of
the supercharge. Nevertheless, our model is different and provides more
realistic generalization of the conventional SUSY.
The upshot of this generalization shows that, 
in addition to the zero-energy state, 
all the midgap states, including finite energy ones, are
annihilated by the supercharge $Q_p$.
The wavefunctions of the midgap states thus obtained tend to
vanish with the same period commensurate with $p$:
$\Psi _{0}\approx (\cdot \cdot \cdot
,0,\cdot \cdot \cdot ,0,\cdot \cdot \cdot ,0,\cdot \cdot \cdot )$.
These zeroes cut the original Hamiltonian 
into smaller ones so that the energies of the midgap states
are determined by the eigenvalues of the Hamiltonian within each period.
As a result, the matrix for determining the energies
of midgap states is of size much smaller than
the size of the original Hamiltonian.
This reduction in matrix size
heavily reduces the computation for determining the occurrence of the midgap
states and provides a way to control the occurrence of the midgap states.
As an application, we propose a structure with superlattice in hopping
with period $p$ to control the number of localized electronic states 
occurring at the end of nanowires.
In that case, the number of edge states
is simply $p-1$.

As the period $p$ goes to infinity, the ensemble of configurations
of hopping forms a semi-infinite disorder chain.
This limit has been extensively investigated
during the past\cite{McKenzie}
since Dyson's seminal work\cite{Dyson} in which it
was pointed out that the average density of state (DOS)
is enhanced at zero energy.
From our point of view, this enhancement also reflects
that the system has high probabilities to take
the above-mentioned form for the ground state.
The presence of the boundary breaks
translational invariance. Thus,
unlike the bulk case where the DOS at zero energy
has no spatial dependence,
the enhance DOS at zero
energy for semi-infinite disordered wires has the largest amplitude
near the edge. Even for slight disorders, the effects of
enhanced DOS at zero-energy are still observable.
This offers a possible explanation for many unexpected
zero-bias anomalies observed in tunneling experiments
because, unless extremely carefully controlled, junction qualities are usually rather poor
and disorders can easily set in near the junctions\cite{Cheng}.

Other implications and extensions of our generalized SUSY quantum
mechanics will also be
discussed. In particular, we shall demonstrate that by appropriate
mappings,
the ordinarily recognized
impurity state can be viewed as a disguised midgap state.
Such mapping provides a simple way to construct the impurity
wavefunction and the corresponding energy. In addition to
this application,
possible extension to include the electron-electron interactions
will also be discussed at the end of this paper.

This paper is organized as follows. In Sec.\ref{Theory}, we lay down
the basic tight binding model considered in this work
and illustrate the SUSY quantum mechanics for the bipartite
systems. In Sec.\ref{General}, we generalize the
supercharge and supersymmetric Hamiltonian
to the $p$-partite systems and discuss the
disorder limit. 
We also point it out of
how to engineer the number of edge states
by using a superlattice structure. By applying the SUSY
quantum mechanics, we illustrate in Sec.\ref{Impurity} how an impurity state
can be viewed as a midgap state. In Sec.\ref{conclude},
we conclude and discuss possible generalization to include
electron-electron interactions. Appendices A and B are devoted to
technical details of superalgebra and computation of commutators.

\section{Theoretical Formulation and Supersymmetric Quantum Mechanics}
\label{Theory}
\begin{figure}
\hspace*{-10mm}
\rotatebox{-90}{\includegraphics*[width=60mm]{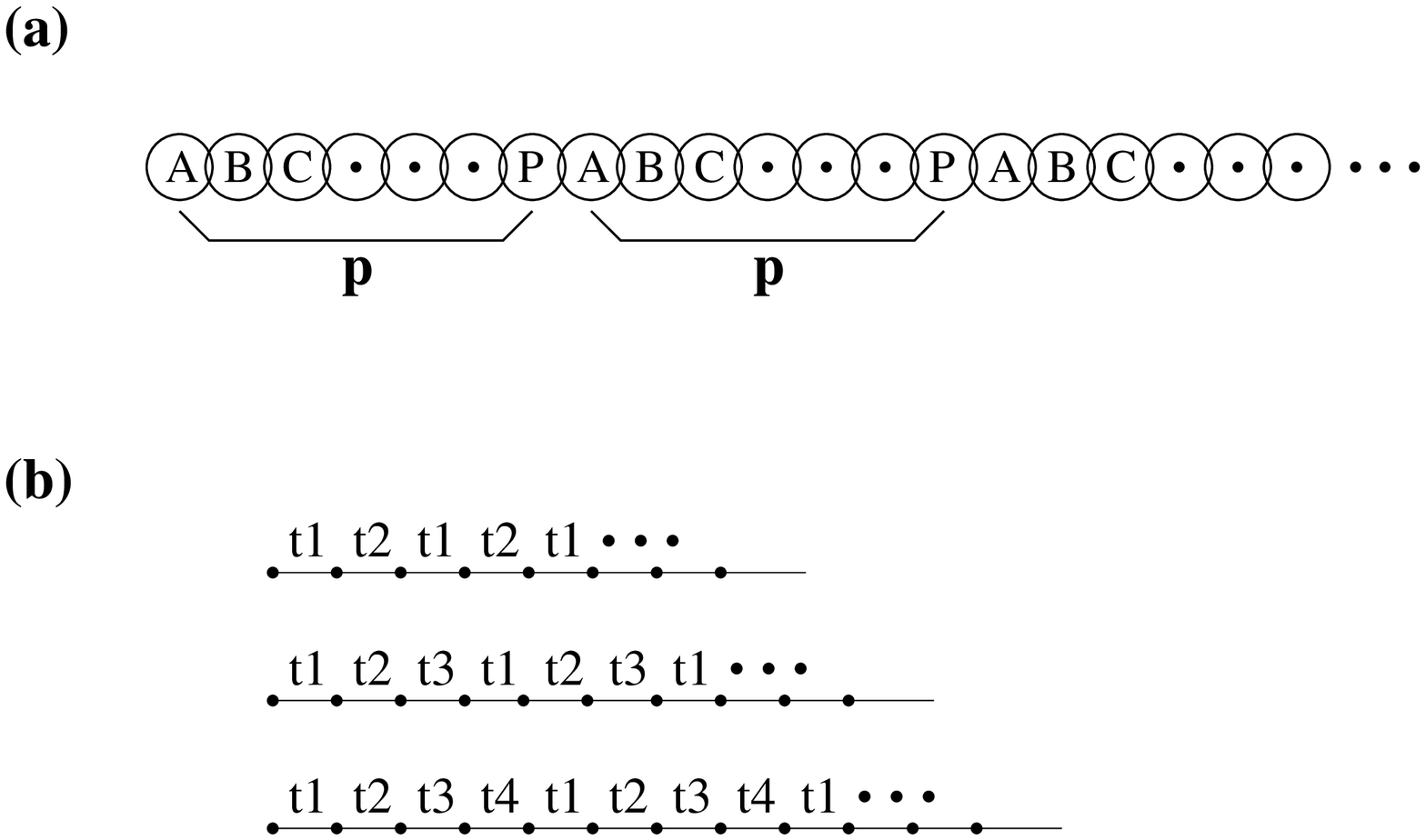}}
\caption{\small Schematic plot of (a) an assembled atomic chain and (b) the
corresponding models with small periods: $p=2$, $3$, and
$4$.\label{chain}}
\end{figure}

We start by considering the 1D atomic chain as illustrated
in Fig.~\ref{chain}(a). This is the most general 1D atomic chain in which
reflection symmetry with respect to the edge point is broken and, consequently,
edge states might arise\cite{mou1}.
In the tight-binding limit, we consider the following Hamiltonian to
model this system:
\begin{equation}
H_{p}=\sum_{i=1}^{\infty }t_{i}c_{i}^{\dagger}
c_{i+1}+h.c.+\upsilon _{i}c_{i}^{\dagger}c_{i}.  \label{H}
\end{equation}
Here the subscript $p$ indicates the period of the lattice and $i$ is
the site index; $t_i$ is the hopping amplitude between site $i$ and
its nearest neighbors, $c_i$ ($c_i^\dagger$) is the electron
annihilation (creation) operator, and $\upsilon_i$ is the local
potential at site $i$. We shall assume that both $t_{i}$ and
$\upsilon_{i}$ are periodic with period $p$, namely $t_{p+i}=t_{i}$
and $\upsilon_{p+i}=\upsilon_{i}$.  In real systems, this Hamiltonian
may correspond to an assembly of $p$ different atoms repeatedly
arranged into a line (see Fig.~\ref{chain}).
For wires composed of atoms of a single species,
$H_p$ may describe systems which exhibit density-wave order. This
includes polyacetylene\cite{Su1}, which has a dimerized structure and
corresponds to $p=2$, and polymers with higher commensurability charge
density waves\cite{Su2}.  In the following, we shall call $p=2$ the
$t_{1}$-$t_{2}$ model, and similarly for models with higher
periods. As mentioned in the introduction, $H_{p}$
may also represent the reduced model of a higher dimensional structure
after partial Fourier transformation.
For example, for a semi-infinite graphite sheet with zig-zag
edge, since the system is translationally invariant along the edge, a
partial Fourier transformation can be applied along this direction,
leading to an effective 1D model. In this case, it is identical to the
$t_{1}$-$t_{2}$ model except that now $t_{1} $ and $t_{2}$ are
$k$-dependent\cite{mou1}: $t_{1}=2 t_0 \cos (\sqrt{3} k_y a/2), t_2=t_0$
, where $a$ is the lattice constant and $k_y$ represents the Fourier mode.
This approach has been successfully applied
to understand the anomalous properties near the edge in carbon
ribbons\cite{mou}. As a final example, we note that the operator
$c_{i}$ in $H_{p}$ needs not be restricted to be the electron
annihilation operator. For example, after applying the Jordan-Wigner
transformation, one can map a 1D quantum XY spin chain to a 1D model
described by $H_{p}$.  Specifically, we have $t_{i}$ replaced by the
exchange coupling for nearest neighbors $J_{i}/2$, and $\upsilon _{i}$
replaced by the local magnetic field $h_{i}$. It is clear from these
examples that $H_{p}$ \ is quite general and captures the physics of
many interesting systems.
\begin{figure}
\hspace*{-10mm}
\rotatebox{-90}{\includegraphics*[width=60mm]{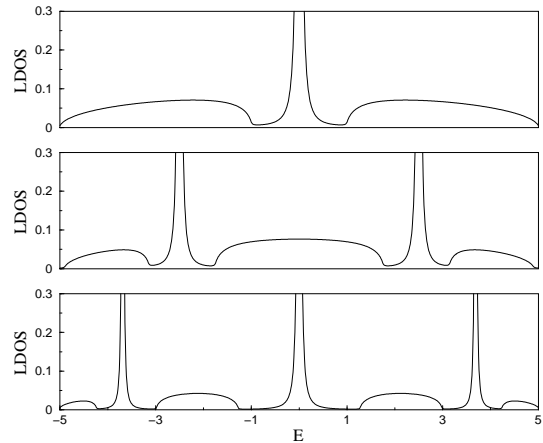}}
\caption{\small  The local density of states at the end point for small periods:
$p=2$, $3$, and $4$. The parameters are carefully chosen so that the midgap states
are manifested. Top: $t_1=2.0$, $t_2=3.0$ ($p=2)$; Middle: $t_1=2.5$, $t_2=1.8$,
$t_3=3.0$ ($p=3$); Bottom: $t_1=2.4$, $t_2=2.8$, $t_3=1.3$, $t_4=3.4$ ($p=4$).
All potential $\upsilon _{i}=0$, and we have included a lifetime $\delta
=0.02$.{\label{LDOS}}}
\end{figure}

To investigate the behavior of $H_{p}$ near the edge, as a first step,
we calculate the local density of states at the end point using the
generalized method of image developed in Ref.~3.
Fig.~\ref{LDOS} shows typical local density of
states at the end point for small periods.
The parameters are carefully chosen so that all possible midgap states are
present. In particular, we have set $\upsilon _{i}=0$, which amounts
to choosing the energy zero as the origin. These results show that midgap
states are indeed the most prominent features at the end point. To
understand how the midgap states arise, we first investigate the
$t_{1}$-$t_{2}$ model with $\upsilon _{i}=0$ in details. In this case,
since the lattice is bipartite, it is convenient to distinguish the
amplitudes at the odd and the even sites by writing the wavefunction
as $\Psi=(\phi_{o}$, $\phi_{e})$. The Hamiltonian then becomes
\begin{equation}
H_{2}=\left(
\begin{array}{cc}
{\bf 0} & {\bf A} \\
{\bf A}^{^{\dagger }} & {\bf 0}
\end{array}
\right) . \label{H2}
\end{equation}
Here ${\bf 0}$ is the null matrix and ${\bf A}$ is a non-Hermitian matrix
\begin{equation}
{\bf A=}\left(
\begin{array}{cccc}
t_{1} & 0 & 0 & \cdot \cdot \\
t_{2} & t_{1} & 0 & \cdot \cdot \\
0 & t_{2} & t_{1} & \cdot \cdot \\
\cdot & \cdot & \cdot & \cdot \cdot
\end{array}
\right) \, . \label{A}
\end{equation}
It is interesting to note that the adjoint of ${\bf A}$ satisfies
\begin{equation}
{\bf A}^{^{\dagger }}
={\bf \varepsilon A\varepsilon }
\quad \mbox{with} \quad
{\bf \varepsilon =}\left(
\begin{array}{cccc}
0 & 0 & \cdot \cdot & 1 \\
0 & \cdot \cdot & 1 & 0 \\
\cdot \cdot & \cdot \cdot & \cdot \cdot & \cdot \cdot \\
1 & 0 & \cdot \cdot & 0
\end{array}
\right) \, .  \label{reflection}
\end{equation}
Here the operator ${\bf \varepsilon }$ effectively reflects the wavefunction
with respect to the mid point of the lattice.

In the case of infinite chains, it is not hard to check that the corresponding matrices
${\bf A}$ and ${\bf A}^{^{\dagger }}$ commute with each other and
hence can be diagonalized simultaneously in Fourier space.
For semi-infinite chains, however, ${\bf A}$ and ${\bf A}^{^{\dagger }}$
do not commute and the spectrum of
the $t_{1}$-$t_{2}$ model can be best understood
in terms of the supersymmetric quantum
mechanics\cite{susy1}. For this purpose, we first identify $H_{2}$ as
the {\it supercharge} $Q_{2}$, which connects even and odd sites. The
block-diagonal matrix $(H_{2})^{2}$ ($\equiv H^S $) is then identified as
(up to a factor of two) the corresponding supersymmetric Hamiltonian,
whose diagonal blocks $H_{o}^{S}\equiv{\bf AA}^{^{\dagger }}$ and
$H_{e}^{S}\equiv{\bf A}^{^{\dagger }}{\bf A}$ are, respectively, the
effective Hamiltonians for the odd and the even sites.
Note that because ${\bf A}$ and ${\bf A}^{^{\dagger }}$ do
not commute, $H_{o}^{S} \neq H_{e}^{S}$. We will show below
that the difference between $H_{o}^{S}$ and $H_{e}^{S}$
is the origin of the midgap states.
Obviously, $H^S$ is positive definite with the possibility
when its spectrum touches zero.
When the later happens, the
ground state energy of $H^S$ vanishes, the ground state
wavefunctions $\phi _{e}$ and $\phi
_{o}$ [$\Psi_0 =(\phi_{o}$, $\phi_{e})$ ]
have to be the zero-energy eigenfunction of ${\bf A}$ and ${\bf A}%
^{^{\dagger }}$, {\it ie.}~${\bf A}\phi _{e}=0$ and
${\bf A}^{^{\dagger }}\phi _{o}=0$. In other words, the supercharge
annihilates the ground state wavefunction $
\Psi_{0}$
\begin{equation}
Q_{2}\Psi _{0}=H_{2}\Psi _{0}=0 \, . \label{super}
\end{equation}
Clearly, in this case, the system has good supersymmetry because
the ground state is invariant under ``rotation" between even and odd
sites:
\begin{equation}
e^{i \theta Q_2 } \Psi_0 = \Psi_0, \label{goodsuper}
\end{equation}
where $\theta$ is any real number.
The non-Hermiticity of ${\bf A}$ and ${\bf A}^{^{\dagger }}$ implies that
forward and backward hopping amplitudes between two sites are different, and
hence the eigenfunctions have to either grow or decay from the end point.
Obviously, because of the relation
${\bf A}^{^{\dagger }}={\bf \varepsilon A\varepsilon }$,
any non-trivial eigenfunctions
satisfy $\phi _{e}={\bf \varepsilon}\phi_{o}$.
Therefore, if $\phi_{o}$ decays from the edge, $\phi_{e}$
must grow from the edge (vice versa).
For semi-infinite chains, only the even sites are connected
with the hard-wall boundary point. Thus
$\phi_{e}$ is forced to vanish while
$\phi_{o}$ decays into the bulk, so that
$\Psi_0 =(\phi_{o}, 0)$. Note that the other possible
state $\Psi_0 =(0, \phi_{e})$ resides on the other end of the chain
and is pushed to infinity. Therefore, overall speaking,
there is only half a chance for the existence of
the ground state $(\phi_{o}, 0)$.
This also reflects in the hopping strength difference.
Indeed, we find that $\phi_{o}$ decays only when
$t_{1}<t_{2}$. In this case, $H_{o}^{S}$ has a non-trivial zero energy
eigenfunction, while $H_{e}^{S}$ does not. Therefore, the system has
good supersymmetry with {\it the ground
state }$\Psi _{0}${\it \ being a localized state.} For finite energies,
however, $\phi _{e}$ and $\phi _{o}$ need not be eigenfunctions of ${\bf A%
}$ and ${\bf A}^{^{\dagger }}$.
Nevertheless, the supersymmetry allows a simple
and elegant way to find the whole spectrum for
the case $p=2$.
This is because $H_{e}^{S}$ has the exact
form as $H_{1}$ ($p=1)$ with $t_{i}=t_{1}t_{2}$ and $\upsilon
_{i}=(t_{1}^{2}+t_{2}^{2})$. Since this
is just the ordinary uniform hopping model, one can easily
write down the eigenstate: $\phi
_{e}(n)=\sin 2nk$. The wavefunction at odd site
can be then found by using the supercharge operator. We find that
$\phi _{o}={\bf A}\phi _{e}/E$ with $E$ being
the spectrum of $H_{2}$ which satisfies
$E^{2}=t_{1}^{2}+t_{2}^{2}+2t_{1}t_{2}\cos 2k$. Since
$E^{2}\geq(t_{1}-t_{2})^{2}$, an energy gap opens up around $E=0$
when $t_{1}\neq t_{2}$. In the case of $t_{1}<t_{2}$, the ground state
$\Psi_{0}$ then arises as a midgap state. Note that $H_{o}^{S}$ is almost
identical to $H_{e}^{S}$ except for the potential energy
$\upsilon_{1}=t_{1}^{2}$ at the end point; the deficit energy $t_{2}^{2}$
is entirely due to the missing bond cut off by the boundary. We will
elaborate on this in Sec.\ref{Impurity}.

\begin{figure}
\hspace*{-10mm}
\rotatebox{-90}{\includegraphics*[width=60mm]{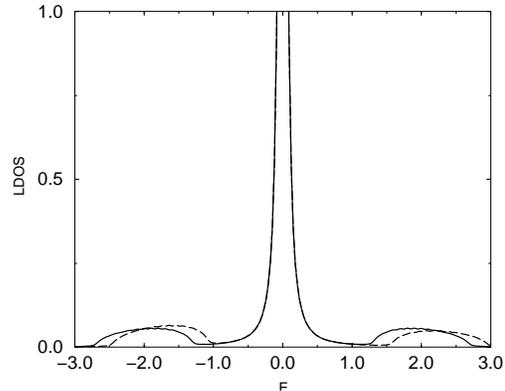}}
\caption{\small The effect of the potential is to break the particle-hole
symmetry so that two side bands are distorted.
However, the midgap state is not changed if $\upsilon_o=0$. The parameters
are: $t_1 =0.7$ and $t_2=2.0$.
For solid line, $\upsilon_o =0, \upsilon_e=0$, while for dash line
$\upsilon_o =0, \upsilon_e=0.5$.\label{ph}}
\end{figure}

We now address the effects of the potential $\upsilon_i$.
For $p=2$, it is convenient to denote the potentials over the even sites
$\upsilon_{e}$ and the odd sites $\upsilon_{o}$. This decomposition,
however, renders the particle-hole symmetry invalid at the level of the
supercharge $H_{2}$.
Nonetheless, the spectrum ($E$) of $H_{2}$ can be mapped
to the original spectrum of $H^S$ with $\upsilon_i=0$ ($\equiv E^0_S$).
For $E^0_{S}\neq 0$ this mapping is
given by $E^0_{S}=(E-\upsilon _{e})(E-\upsilon_{o})$, while for $E^0_{S}=0$,
since $\phi _{e}=0$ still holds, one has $E=\upsilon _{o}$. 
Hence even though the physical spectrum $E$ may have no particle-hole
symmetry, after appropriate transformations, the symmetric structure can
be restored in $E^0_{S}$; in particular, the midgap state
survives as clearly demonstrated in Fig.~\ref{ph}.
For higher periods, the same manipulations as above
can lead to similar conclusions. Therefore, unless explicitly needed, we
shall ignore $\upsilon _{i}$
in the following.

%We note in passing that a uniform potential is equivalent
%to a chemical potential. In the 1D quantum XY spin chain, a uniform magnetic
%field $h$ plays the role of the chemical potential and the averaged particle
%density $\langle c_{i}^{\dagger}c_{i}\rangle $ corresponds to the local
%magnetization $M^{i}_{z}$.
%For $h$ large enough so that the midgap states become
%populated, a large magnetization is then induced near the edge.
%This is illustrated in Fig.~\ref{ph}.
%\begin{figure}
%\hspace*{-10mm}
%\rotatebox{-90}{\includegraphics*[width=60mm]{fig3.eps}}
%\caption{\small A small external magnetic field can induce large
%edge magnetization in a 1D
%quantum XY model.
%It is conceivable that such an effect may be used for probing magnetism.}
%\end{figure}

Let us now apply the supersymmetric method to
semi-infinite superconductors\cite{Legget}.
After partial Fourier transformation
along the interface, the problem becomes 1D superconductors with
an end point. In this case, it is convenient to write
the wavefunction by $\Psi =(u, v)$ with $u=(u_1,u_2,u_3,\cdots)$
being particle-like
and $v=(v_1,v_2,v_3,\cdots)$ being hole-like wavefunctions. The reduced
mean-field (BCS) Hamiltonian is Dirac-like\cite{susy1}
and can be generally written as
\begin{equation}
H_{BCS}=\left(
\begin{array}{cc}
\mathbf{M} & \mathbf{Q} \\
\mathbf{Q} & \mathbf{-M}
\end{array}
\right),
\label{BSC}
\end{equation}
where $\mathbf{M}$ corresponds to the reduced 1D Hamiltonian
for particles and $\mathbf{Q}$ is essentially the pairing
potential. One can also rewrite $H_{BSC}
=\mathbf{M\otimes \sigma }_{z}+\mathbf{Q\otimes \sigma }_{x}$,
and treat this problem as a spin in the
``magnetic field" $(\mathbf{Q},0,\mathbf{M})$
pointing in the $x-z$ plane.
This analogy suggests that it is possible to rotate the
magnetic field to the $x-y$ plane.
Indeed, this can be achieved by a rotation of
$2 \pi /3$ with respect to the axis $(1,1,1)$.
The transformation matrix\cite{Legget} is
\begin{equation}
U=\frac{1}{\sqrt{2}}\left(
\begin{array}{cc}
\mathbf{1} & \mathbf{1} \\
\mathbf{i} & \mathbf{-i}
\end{array}
\right) ,
\end{equation}
where $\mathbf{1}$ and $\mathbf{i}$ are semi-infinite matrices.
The rotated
Hamiltonian then takes the form of a supercharge like $H_{2}$
\begin{equation}
H_{BCS}^{\prime }=U^{\dagger }H_{BCS}U=\left(
\begin{array}{cc}
\mathbf{0} & \mathbf{M-}i\mathbf{Q} \\
\mathbf{M+}i\mathbf{Q} & \mathbf{0}
\end{array}
\right) .  \label{BSC1}
\end{equation}
The wavefunction is rotated accordingly: $\Psi ^{\prime
}=U^{\dagger }\Psi $. Therefore, in the supersymmetric form, particles and holes are
mixed. As an illustration, we consider
the mean-field Hamiltonian for $d$-wave superconductors
\begin{eqnarray}
H_R &=&-\!\!\! \sum_{\langle ij \rangle,\sigma}
t_0 c_{i\sigma }^\dag c_{j\sigma}^{}
+ \Delta _{ij}
(c_{i\uparrow }c_{j\downarrow}-c_{i\downarrow}c_{j\uparrow })
+{\rm h.c.} \,,  \label{H_NS}
\end{eqnarray}
where $\langle ij \rangle$ denotes the nearest-neighbor bonds,
$t_0$ and $\Delta_{ij}$ are, respectively, the corresponding hopping and
$d$-wave pairing amplitudes.
For the $(1,1,0)$ interface, after Fourier transformation
along the interface (which is taken to be the $y$ direction), we obtain\cite{mou1}
\begin{equation}
\mathbf{A}=\mathbf{M-iQ}=\left(
\begin{array}{cccc}
-\mu  & -t+d  & 0 & \cdot \cdot  \\
-t-d  & -\mu  & -t+d  & \cdot \cdot  \\
0 & -t-d  & -\mu  & \cdot \cdot  \\
\cdot  & \cdot  & \cdot  & \cdot \cdot
\end{array}
\right),
\end{equation}
where $\mu$ is the chemical potential,
$t=2t_{0}\cos (k_{y}a/\sqrt{2})$ and $d=2\Delta _{0}\sin (k_{y}a/\sqrt{%
2})$. The model for tight-binding $d$-wave superconductors
is unique in the sense that the
non-Hermiticity of $\mathbf{A}$ and ${\bf A}^{^{\dagger }}$
can be removed by a gauge transformation.
For this purpose, we write $t-d=\tilde{t} e^{g}$ and $t+d=\tilde{t} e^{-g}$
with $\tilde{t}=\sqrt{t^2 -d^2}$ and $e^{2g}=(t-d)/(t+d)$. The eigenfuction
$\Psi_E$ of $\mathbf{A}$ is thus the gauge transformation
of the eigenfuction $\phi_E$ of
\begin{equation}
\mathbf{A}_{0}=\left(
\begin{array}{cccc}
-\mu  & -\tilde{t} & 0 & \cdot \cdot  \\
-\tilde{t} & -\mu  & -\tilde{t} & \cdot \cdot  \\
0 & -\tilde{t} & -\mu  & \cdot \cdot  \\
\cdot  & \cdot  & \cdot  & \cdot \cdot
\end{array}
\right).
\end{equation}
Specifically, we obtain $\Psi_E(n) = e^{-ng} \phi_E$
(for ${\bf A}^{^{\dagger }}$,
one obtains $\Psi_E(n) = e^{ng} \phi_E$). Furthermore,
$\Psi_E$ and $\phi_E$ have the same eigenvalue $E$.
This implies that
the calculation of the zero mode
is related to the spectrum of $\mathbf{A}_{0}$.
If we restrict our discussion to the propagating modes, 
the spectrum of $\mathbf{A}_{0}$ is simply
the ordinary cosine band.
We find that when $-2\sqrt{t^2-d^2} < \mu < 2\sqrt{t^2 -d^2}$
is satisfied, $\mathbf{A}$ and ${\bf A}^{^{\dagger }}$ can, 
respectively, support
zero-modes of the form $(0,e^{-ng} \phi_{E=0})$ and
$(e^{ng} \phi_{E=0},0)$.
In this case, the ground state of $H^S$ is selected by the sign of $g$,
and thus the zero-energy midgap state is given by
\begin{eqnarray}
u(n) &=&\frac{1}{\sqrt{2}}{\left( \frac{t-|d|}{t+|d|}\right) }%
^{n/2}\sin(k_{F}na), \\
v(n) &=&{\rm sign}(d)\frac{i}{\sqrt{2}}{\left( \frac{t-|d|}{t+|d|}\right) }%
^{n/2}\sin(k_{F}na).
\end{eqnarray}
Here $k_{F}$ is determined by $-2\sqrt{t^{2}-d^{2}}\cos(k_{F}a)=\mu $
and depends on $k_{y}$; therefore, even though the midgap states
for different $k_{y}$'s
have the same zero energy, their wavefunctions have $k_{y}$ dependence. 
Note that for demonstration, we have only considered the
case when $k_F$ is real. Complete
solutions, however, require to include the situation 
when $k_F$ is complex\cite{hsiuhau}. In both cases,
the supersymmetry structure enables one to write down 
the explicit form of the zero-energy
mode near the interface $(1,1,0)$.

\section{Generalized Supercharge and its consequences}
\label{General}
We now generalize the above results to higher periods $p\geq 3$.
It is useful to
decompose the wavefunction
as $\Psi =(\phi _{1}$, $\phi _{2}$,$\cdots $,$\phi _{p})$, where
$\phi _{n}$ denotes the sub-wavefunction formed by
$\{\Psi (kp+n); k=0,1,2,\dots\}$. The Hamiltonian is then cast in
the form
\begin{equation}
H_{p}=\left(
\begin{array}{ccccc}
{\bf 0} & {\bf A}_{12} & {\bf 0} & \cdot \cdot  & {\bf A}_{1p} \\
{\bf A}_{12}^{^{^{\dagger }}} & {\bf 0} & {\bf A}_{23} & \cdot \cdot  & {\bf 0}
\\
{\bf 0} & {\bf A}_{23}^{^{^{\dagger }}} & {\bf 0} & \cdot \cdot  & \cdot  \\
\cdot \cdot  & \cdot \cdot  & \cdot \cdot  & \cdot \cdot  & {\bf A}_{p-1,p} \\
{\bf A}_{1p}^{^{^{\dagger }}} & {\bf 0} & \cdot \cdot  & {\bf A}%
_{p-1,p}^{^{^{\dagger }}} & {\bf 0}
\end{array}
\right) .
\end{equation}
Again, here ${\bf 0}$ and ${\bf A}_{nm}$ are block matrices; for all $n \neq p$,
${\bf A}_{nm}=t_{n}{\bf 1}$ are diagonal, while for $(m,n)=(1,p)$
\begin{equation}
{\bf A}_{1p}=\left(
\begin{array}{cccc}
0 & 0 & 0 & \cdot \cdot  \\
t_{p} & 0 & 0 & \cdot \cdot  \\
0 & t_{p} & 0 & \cdot \cdot  \\
\cdot  & \cdot  & \cdot  & \cdot \cdot
\end{array}
\right) . \label{Ap}
\end{equation}
To understand what happens for the semi-infinite chain, it
is useful to start from the infinite chain with Hamiltonian
$H^{\infty}_p$. In this case,
$H^{\infty}_p$ also takes the same form
except that ${\bf A}_{nm}$ are further extended to $i=- \infty$.
If we remove the hopping strength $t_n$ and combine the remaining
${\bf A}_{nm}$ with ${\bf A}_{nm}^{^{^{\dagger }}}$ 
into ${\bf Q}_{nm}$
for all $m$ and $n$ pairs, ${\bf Q}_{nm}$
form a superalgebra if modulo $p$ is performed 
(see Appendix A for mathematical details).
The energy bands of $H^{\infty}_p$ are determined by
\begin{widetext}
\begin{equation}
P(E,k) = \det \left(
\begin{array}{ccccc}
E & -t_{1}e^{ik} & 0 & \cdot \cdot  & -t_{p}e^{-ik} \\
-t_{1}e^{-ik} & E & -t_{2}e^{ik} & \cdot \cdot  & 0 \\
0 & -t_{2}e^{-ik} & E & \cdot \cdot  & \cdot  \\
\cdot \cdot  & \cdot \cdot  & \cdot \cdot  & \cdot \cdot  & -t_{p-1}e^{ik}
\\
-t_{p}e^{ik} & 0 & \cdot \cdot  & -t_{p-1}e^{-ik} & E
\end{array}
\right) =0, \label{polynomial}
\end{equation}
\end{widetext}
where $E$ is the energy and $k$ is the Fourier mode. 
In general, there have at most $p$ energy bands.
However, since in the polynomial $P(E,k)$,
$(-1)^{p+1} 2 t_{1}t_{2}t_{3}\cdot \cdot \cdot t_p cos(pk)$
is the only term that depends on $k$,
the function $P(E,k) -(-1)^{p+1} 2 t_{1}t_{2}t_{3}\cdot \cdot \cdot
 t_{p} cos(pk)$ maps $p$ bands into one
single band: $2 t_{1}t_{2}t_{3}\cdot \cdot \cdot t_{p} cos(pk)$.
This important observation implies that when $H_p$ = $H^{\infty}_p$,
the operator $H_S \equiv P(H_p,0) -
(-1)^{p+1} 2 t_{1}t_{2}t_{3}\cdot \cdot \cdot t_{p}$
is block-diagonal [there are $p$ blocks with one for each
$\phi_i$, see Eqs.(\ref{block1}) and (\ref{block2})] and
folds the
spectrum of $H^{\infty}_p$ into one single band\cite{para1}.
Therefore, $H_S$ is similar to
the supersymmetry Hamiltonian $H_S$.
Indeed, for $p=2$, we find $H_S=H^2_2 -(t^2_1 + t^2_2)$ 
which is essentially $H^S$.

For infinite chains, $H_S$ is highly symmetric. In fact,
it commutes with all $Q_{mn}$. 
This reflects that it is symmetric
under the permutation of $n$ and $m$ but it is more than
that because any linear combination of $\sum t_{mn} Q_{mn}$
also commutes with $H_S$. For semi-infinite chains, however,
the above symmetry is broken: Not all $Q_{mn}$
commute with $H_S$.
Physically, this is obvious because now $\phi_{p}$ is special
and is the only component 
that connects with the boundary point $i=0$ directly.
As a result, $H_S$ is not completely block-diagonalized.
In fact, because even for the infinite chains, 
$\phi _{p}(n)=\sin(knpa)$ is a solution for the $p$th block 
and it satisfies the hard-wall boundary condition at $i=0$,
the $p$th block is not affected. Therefore, 
there are only two blocks: one for the space formed by $\phi_p$;
the other mixes $\phi_1$,$\phi_2$,..and $\phi_{p-1}$.
This is demonstrated in Eq.(\ref{block1}) where we
denote the block Hamiltonians by $\mathbf{H}^{+}$ and $\mathbf{H}^{-}$.

Clearly, the
$t_1$-$t_2$ model is special because
$\mathbf{H}^{+}$ and $\mathbf{H}^{-}$ are of the same size so that
$H_S$ is completely block-diagonal. This is where
the usual SUSY quantum mechanics applies.
For $p \geq 3$, $\mathbf{H}^{+}$ and $\mathbf{H}^{-}$
are not of the same size, a generalization of SUSY quantum mechanics
is needed.
%As a result, extra states (the midgap states),
%in addition to the zero-energy state, may arise.
First, it is important to see if one can find an operator,
similar to the supercharge $Q_2$, that commutes with $H_S$.
$H_p$ is obviously a solution because
$H_2 = Q_2$. However,
in analogy to the case of $p=2$, 
a second supercharge by collecting all block matrices in $H_p$ that
 connects $\phi _{p}$ to other components can be formed:
\begin{equation}
Q_{p}=\left(
\begin{array}{ccccc}
{\bf 0} & {\bf 0} & {\bf \cdot \cdot } & \cdot \cdot  & {\bf A}_{1p} \\
{\bf 0} & {\bf 0} & {\bf \cdot \cdot } & \cdot \cdot  & {\bf 0} \\
{\bf \cdot \cdot } & \cdot \cdot  & {\bf 0} & \cdot \cdot  & \cdot \cdot  \\
\cdot \cdot  & \cdot \cdot  & \cdot \cdot  & \cdot \cdot  & {\bf A}_{p-1,p} \\
{\bf A}_{1p}^{^{^{\dagger }}} & {\bf 0} & \cdot \cdot  & {\bf A}%
_{p-1,p}^{^{^{\dagger }}} & {\bf 0}
\end{array}
\right) \text{ \ for }p\geq 3\text{.}  \label{super1}
\end{equation}
Note that the above defintion can also include
$p=2$. In that case, one squeezes the block
${\bf A}_{1p}$ into ${\bf A}_{12}$ to obtain the
form of $H_2$ in Eq.(\ref{H2}). Because ${\bf A}_{12}=t_{1}{\bf 1}$
and ${\bf A}_{1p}$ is given by Eq.(\ref{Ap}) with $t_p$ being replaced
by $t_2$, adding ${\bf A}_{12}$ and ${\bf A}_{1p}$ reproduces
${\bf A}$ defined in Eq.(\ref{A}) precisely. Hence Eq.(\ref{super1})
can be regarded as an "analytical continuation" of $Q_{2}$
to $p \geq 3$. Furthermore, as shown in
Appendix B, $[H_S,Q_{p}]=0$ is satisfied, thus
$Q_{p}$ provides a faithful generalization of $Q_{2}$.
It coincides with $H_p$ only
when $p=2$. Note that from Appendix B, one can actually see
that out of $Q_{mn}$ contained in $H_p$,
$Q_p$ and $H_p$ (and their linear combinations) are 
the only two generators that commute with $H_S$ for general $p$.

Applying the condition that the supercharge annihilates midgap states
$\Psi _{0}$, one finds that
\begin{equation}
\phi _{p}=0\quad \text{ and }\quad
{\bf A}_{1p}^{^{^{\dagger }}}\phi _{1}+{\bf A}%
_{p-1,p}^{^{^{\dagger }}}\phi _{p-1}=0.  \label{phip}
\end{equation}
$\phi _{p}=0$ \ implies that the wavefunction has the form $%
\Psi _{0}\approx (\cdot \cdot \cdot ,0,\cdot \cdot \cdot ,0,\cdot \cdot
\cdot ,0,\cdot \cdot \cdot )$ as pointed out earlier\cite{complete}, while the
second condition relates $\phi _{p-1}$ to $\phi _{1}$. It is important to
realize that because $Q_{p}$ no longer coincides with $H_{p}$ for $p\geq 3$
, $Q_{p}$ alone does not determine the energies and
wavefunctions of the midgap states. Instead, because $Q_{p}\Psi _{0}=0$, the
operator $H_{p}-Q_{p}$ determines the energies and further provides
relations between $\phi _{2}$,$\cdots$,$\phi _{p-1}$ and $\phi _{1}$.
This analysis shows that the
energies $E_{m}$ of midgap states can be different from zero and must satisfy
\begin{equation}
\det \left(
\begin{array}{ccccc}
-E_{m} & t_{1} & 0 & \cdot \cdot  & 0 \\
t_{1} & -E_{m} & t_{2} & \cdot \cdot  & 0 \\
0 & t_{2} & -E_{m} & \cdot \cdot  & \cdot  \\
\cdot \cdot  & \cdot \cdot  & \cdot \cdot  & \cdot \cdot  & t_{p-2} \\
0 & 0 & \cdot \cdot  & t_{p-2} & -E_{m}
\end{array}
\right) =0.  \label{mid}
\end{equation}
Therefore, there are at most $p-1$ midgap states. 
To stabilize the midgap states, one further requires
$\Psi _{0}$ to decay away
from the edge. In the case of $p=2$, this results in
the condition $t_1 < t_2$.
For $p=3$,
one first obtains from Eq.(\ref{mid}) $E_{m}$ $=$ $\pm t_{1}$ and $
\phi _{1}=\pm \phi _{2}$, which, when combined with Eq.(\ref{phip}), results
in $\Psi (3k+n)=\pm t_{3}/t_{2}\Psi (3k-3+n)$.
Thus midgap states exit only when $t_{2}<t_{3}$. In general, one
needs to relate $\psi_{p-1}$ to $\psi_1$. 
This further reduces the matrix in Eq.(\ref{mid}) and by
defining the $(p-2) \times (p-2)$ matrix
\begin{equation}
\mathbf{h}=\left(
\begin{array}{ccccc}
-E_{m} & t_{1} & 0 & \cdot \cdot  & 0 \\
t_{1} & -E_{m} & t_{2} & \cdot \cdot  & 0 \\
0 & t_{2} & -E_{m} & \cdot \cdot  & \cdot  \\
\cdot \cdot  & \cdot \cdot  & \cdot \cdot  & \cdot \cdot  & t_{p-3} \\
0 & 0 & \cdot \cdot  & t_{p-3} & -E_{m}
\end{array}
\right),  \label{mid1}
\end{equation}
we find that $\psi_1 = - t_{p-2} \mathbf{h}^{-1}_{1,p-2} \psi_{p-1}$.
When combined with Eq.(\ref{phip}), we obtain
$\psi_1 =  t_{p-2} t_p /t_{p-1} \mathbf{h}^{-1}_{1,p-2} \psi_{p+1}$.
Hence the midgap state with energy $E_m$ exists only when 
$t_{p-2} t_p \mathbf{h}^{-1}_{1,p-2} > t_{p-1}$\cite{index}.
Note that for higher periods, commensurate structures may appear 
in sublattices. These structures resemble the SUSY structures
in lower periods. For example, when $p=4$,
there are at most three midgap states at $E_{m}$ $=0$, $\pm
\sqrt{t_{1}^{2}+t_{2}^{2}}$. In this case,
the Hamiltonian $H^2_4$ is already block diagonal in even
and odd sites.
For even sites, $H^2_4$
is period of two and belongs to $H_{2}$ with $t_{1}^{\prime }=t_{1}t_{4}$, $%
t_{2}^{\prime }=t_{2}t_{3}$, $\mu _{1}^{\prime }=t_{1}^{2}+t_{2}^{2}$, and $%
\mu _{2}^{\prime }=t_{3}^{2}+t_{4}^{2}$ with the midgap energy given
by $E_{m}^{\prime }=\mu _{1}^{\prime }=$ $t_{1}^{2}+t_{2}^{2}$. It is clear
that $E_{m}^{\prime }$ is precisely the square of $E_{m}=\pm \sqrt{%
t_{1}^{2}+t_{2}^{2}}$, and thus $\pm \sqrt{t_{1}^{2}+t_{2}^{2}}$ are
the midgap states resulting from the supersymmetry within the even
sublattice. The remaining midgap state at $E_{m}=0$ originates from the
same supersymmetric structure as that for $p=2$ 
between even and odd sites. Hence $\phi_e$ vanishes for $E_m=0$.
Obviously, the same effects may occur for any $p$ that is not
a prime number.

As an application, we point out that a simple way to engineer localized
edge states in a nanowire (such as carbon nanotubes) is to introduce
impurity-bonds periodically and form a semi-infinite superlattice structure.
In this case, one has $t_1= t_2 = \cdots = t_{p-1} \equiv t$ while
$t_p \equiv t'$. 
It is known that in the infinite case, there are $p$ energy bands.
%Far from the edge point, the nanowire is not affected
%by the edge, so the spectrum of delocalized states are exactly the same
%as that of the bulk. Therefore, the impurity-bond $t'$ folds the original
%Brillouin zone $p-1$ times and results in $p$ mini-bands. 
%This is a well-known
%result for the bulk superlattice structure. What is interesting,
However, when it becomes semi-infinite,
$p-1$ localized edge states may arise simultaneously
and
they are located exactly in the $p-1$ gaps among these energy bands.
The proposed superlattice structure thus provides a convenient way to
control the number of edge states.
The above result can be quite easily derived in
our formulation. First of all, Eq.(\ref{mid}) implies that 
the energies of edge modes
are exactly the energies of a finite atomic chain with uniform hopping $t$.
In other words, $E= 2t \cos(ka)$
with $ka=m \pi/p$, $m=1,2,\cdots,p-1$. Note that these $k$'s occur exactly at
the zone boundaries of the energy bands, 
hence they appear within the energy gaps.
Eq.(\ref{phip}) further implies that for any edge mode, their
wavefunction satisfies $\Psi(p+1)=-t/t' \Psi(p-1)$. Since
$\Psi(1)$ and $\Psi(p-1)$ have the same amplitude, we get
$|\Psi(p+1)|=|t/t'| |\Psi(1)|$. Therefore, as long as
$|t/t'|<1$, all midgap wavefunctions
decay and hence the $p-1$ midgap states appear at the same time.

\begin{figure}
\hspace*{-10mm}
\rotatebox{-90}{\includegraphics*[width=60mm]{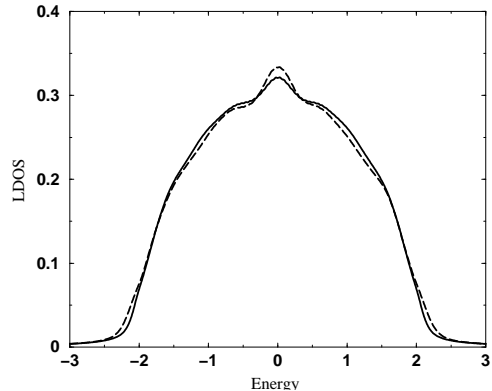}}
\caption{\small
Averaged local density of states over 10000 samples
at the end point for semi-infinite
wires with disorders near the edge point. Only $5$ lattice
points are imposed with disorders. Here random bonds and potentials
are imposed on a uniform hopping model with
$t=1.0$, and for solid line,
the amplitudes of disorders are $\delta t =0.2$ and $\delta \upsilon=0.2$; while
for the dash line, the amplitudes of disorders are $\delta t =0.3$ and
$\delta \upsilon=0.3$. One sees that slight disorders can induce a peak-like
structure at zero energy.\label{disordered}}
\end{figure}

Another example is to consider the limit
when $p$ goes to infinity. Since for any given configuration
of $\{ t_i, i=1,2,3 \dots \}$, $t_i$ ($1 \leq i \leq p=\infty$) is allowed to be any number; effectively,
the wire is a disordered semi-infinite wires.
It is well known that the density of
states at zero-energy become enhanced\cite{Dyson,McKenzie}
in the disordered wires.
In the following, we will find
that such enhancement can be also understood from the
above point of view. Essentially, this is because for any given
$t_i$ configuration, the system has high probability to
settle into the form of the ground
state wavefunction discussed above.
First, because the boundary breaks the translational
invariance, if one decomposes 
the wavefunction
as $\Psi=(\phi_{o}$, $\phi_{e})$,
it is still true that only
the even sites are connected to the hard-wall boundary point.
In this case, for any set of $\{t_{i}\}$,
the non-Hermitian matrix ${\bf A}$ that enters Eq.(\ref{H2})
is given by
\begin{equation}
{\bf A=}\left(
\begin{array}{cccc}
t_{1} & 0 & 0 & \cdot \cdot  \\
t_{2} & t_{3} & 0 & \cdot \cdot  \\
0 & t_{4} & t_{5} & \cdot \cdot  \\
\cdot  & \cdot  & \cdot  & \cdot \cdot
\end{array}
\right) \, .   \label{disorder}
\end{equation}
For the zero-energy state, because ${\bf A} \phi_e =0$,
it is easy to see that $\phi _{e}=0$; while the wavefunction
at odd sites is determined by ${\bf A}^{^{\dagger }} \phi_o=0$.
If we set $\Psi _{0}^{1}=1$,
the wavefunction at site $2N+1$ is given by $\Psi _{0}^{2N+1}\approx
(t_{1}t_{3}t_{5}...t_{2N-1})/(t_{2}t_{4}t_{6}...t_{2N})$. Let
$x_i\equiv|t_{2i-1}/t_{2i}|$, then $\ln |\Psi _{0}^{2N+1}|\approx
\sum_{i=1}^{N}\ln x_{i}$. Clearly, because
$\ln |\Psi _{0}^{2N+1}| - \ln |\Psi _{0}^{2N-1}| \approx \ln x_N$,
the logarithm of the wavefunction at odd sites behaves effectively as a random walker.
Since a random walker has large probability to go to $\pm \infty$,
$\Psi _{0}$
has high probability of decaying to zero far from the edge and
becomes a localized zero-energy mode. This analogy leads to
$\langle (\ln |\Psi _{0}^{2N+1}|)^2 \rangle = \sqrt{N} \sigma$ with $\sigma$
being the standard deviation of $\ln x$. In other words,
$|\Psi _{0}^{2N+1}| \sim e^{\pm \sigma \sqrt{N}}$ where $\pm$ correspond
to states localized at either ends, respectively.
We emphasize that this analysis indicates that
{\em only the standard deviation of $\ln x$ is relevant}
and there is no need for the assumption of Gaussian-type randomness,
which is often invoked in previous works.
Furthermore, the random-walk nature makes the zero-energy peak
much more easily formed. This is illustrated in Fig.~\ref{disordered} where
we show that even slight disorders near the edge
may induce features resembling zero-bias peaks in tunneling measurements.
In this case,
the zero-energy state tends to decay from the edge
but will not become localized.
Instead, after joining the non-disorder bulk region,
it becomes a resonant state\cite{Lieber}.
Such phenomena may have already been seen in experiments\cite{Cheng}.

\section{Impurity states as midgap states}
\label{Impurity}

\begin{figure}
\hspace*{-10mm}
\rotatebox{-90}{\includegraphics*[width=60mm]{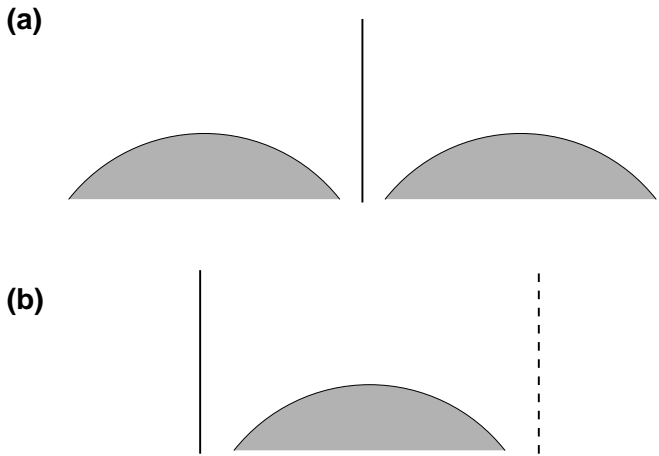}}
\caption{\small Schematic plots of (a) the spectrum with a midgap state and
(b) the spectrum with an impurity state, which may appear at
the position of the solid line or the dashed line. Clearly,
(b) may be viewed as the square of (a) (see text for details), i.e.,
if one folds (a) with respect to the midgap state, one gets (b).
\label{impurity}
}
\end{figure}

In this section, we demonstrate that in the supersymmetric approach
the ordinarily recognized
impurity states can be viewed as disguised midgap states.
Let us consider the Goodwin model for the surface state\cite{Goodwin}.
In this model, it was proposed that the surface state arises
because the potential suddenly changes near the surface or the edge. In the
tight binding limit, the Hamiltonian is given by\cite{Goodwin}
\begin{equation}
H_{G}=\sum_{i=1}^{\infty }t c_{i}^{^{\dagger
}}c_{i+1}+h.c.+U c_{1}^{\dagger}c_{1}.  \label{HG}
\end{equation}
In other words, there is an impurity potential localized at the first site.
It is commonly recognized that under appropriate condition,
an impurity state (in this case, it is the Goodwin edge state.)
may arise and exhibit as an isolated line
in the spectrum as illustrated in Fig.~\ref{impurity}(b). Clearly,
we see from Fig.~\ref{impurity}, the spectrum with an impurity state
is essentially the square of spectrum shown in Fig.~\ref{impurity}(a), i.e.,
the spectrum with a midgap state.

To establish the relation described above, we go back to previous analysis on
the $t_{1}$-$t_{2}$ model. As is obvious, the spectrum of $H^{S}$ ($\equiv
(H_{2})^{2}$) is the square of that for $H_{2}$, fulfilling the relation
indicated in Fig.~\ref{impurity}. It is hence useful and more transparent if
we explicitly write down $H^{S}$
\begin{equation}
H^{S}=\left(
\begin{array}{cc}
\mathbf{AA}^{\dagger } & \mathbf{0} \\
\mathbf{0} & \mathbf{A}^{^{\dagger }}\mathbf{A}
\end{array}
\right) ,
\end{equation}
where
\begin{equation}
H_{o}^{S}=\mathbf{AA}^{\dagger }\mathbf{=}\left(
\begin{array}{cccc}
t_{1}^{2} & t_{1}t_{2} & 0 & \cdot \cdot  \\
t_{1}t_{2} & t_{1}^{2}+t_{2}^{2} & t_{1}t_{2} & \cdot \cdot  \\
0 & t_{1}t_{2} & t_{1}^{2}+t_{2}^{2} & \cdot \cdot  \\
\cdot  & \cdot  & \cdot  & \cdot \cdot
\end{array}
\right) \,
\end{equation}
is the effective Hamiltonian for the odd sites and
\begin{equation}
H_{e}^{S}=\mathbf{A}^{\dagger }\mathbf{A=}\left(
\begin{array}{cccc}
t_{1}^{2}+t_{2}^{2} & t_{1}t_{2} & 0 & \cdot \cdot  \\
t_{1}t_{2} & t_{1}^{2}+t_{2}^{2} & t_{1}t_{2} & \cdot \cdot  \\
0 & t_{1}t_{2} & t_{1}^{2}+t_{2}^{2} & \cdot \cdot  \\
\cdot  & \cdot  & \cdot  & \cdot \cdot
\end{array}
\right)
\end{equation}
is the effective Hamiltonian for the even sites.
One sees that $H_{o}^{S}$ and $H_{e}^{S}$ differ by the potential at site $1$
. As mentioned, this is entirely due to the missing bond cut off by the
boundary. On the other hand, even though $H^{S}$ is block-diagonal, it does
not imply even and odd sites are independently from each other. In fact,
they are connected by the supercharge $H_{2}$. The point is that except for
the zero energy which is an eigenvalue of $H_{o}^{S}$,  $H_{o}^{S}$ and $%
H_{e}^{S}$ share the same energy eigenvalue $E^{2}$ with $E$ being the
spectrum of $H_{2}$. The supersymmetric relation between even and odd sites
enables one to solve the Goodwin model $H_{G}$ as follows. One first rewrites
\begin{equation}
H_{o}^{S}=\sum_{i=1}^{\infty }\left[ (t_{1}t_{2})c_{i}^{^{\dagger
}}c_{i+1}+h.c.+(t_{1}^{2}+t_{2}^{2})c_{i}^{^{\dagger }}c_{i}\right]
-t_{2}^{2}c_{1}^{\dagger }c_{1}.
\end{equation}
Clearly, it shows that the effective Hamiltonian for the odd sites ie
equivalent to the Goodwin model with $t=t_{1}t_{2}$, $U=-t_{2}^{2}$, and $%
\mu =-(t_{1}^{2}+t_{2}^{2})$. The wavefunction of the midgap state on odd
sites then become the wavefunction of the impurity state in the Goodwin
model. Furthermore, the energy of the impurity state can be easily found to
be $E_{im}=-(t_{1}^{2}+t_{2}^{2})$. By solving $t_{1}$ and $t_{2}$ in terms
of $t$ and $U$, we find $E_{im}=U+t^{2}/U$. Since the midgap states
exist when $|t_{1}|<|t_{2}|$, the impurity state appears only when $|t|<|U|$%
, in consistent with the standard approach\cite{Goodwin}. The discussion above concerns
with the case $U<0$, hence the impurity energy resides on the left side (the solid line) in
Fig.~\ref{impurity}(b). For $U>0$, the Goodwin model maps to $-H_{o}^{S}$. One
obtains the same expression for the impurity energy $E_{im}=U+t^{2}/U$
except that it resides on the right side (the dashed line)
in Fig.~\ref{impurity}(b).

In addition to the energy of the impurity states, the above analysis also
implies that the entire spectrum is simply $E_{G}=2t\cos ka$. Furthermore, the
supersymmetric relation between even and odd sites enables one to write down
all wavefunctions for the Goodwin model explicitly. This is entirely due to
the fact that the Hamiltonian of the supersymmetric partner to the Goodwin
model is $H_{e}^{S}$ which is a uniform hopping model. We obtain the
wavefunction for the impurity state $\Psi _{G}(n)=(t/U)^{n-1}$, while
for the extended states, when $U<0$, $\Psi _{G}(n)=[\sqrt{|U|}\sin nka+t/\sqrt{%
|U|}\sin (n-1)ka]/E_{k}$ with $E_{k}=\pm \sqrt{|U|+t^{2}/|U|+2t\cos ka}$,
and when $U>0$, $\Psi _{G}(n)=[\sqrt{|U|}\sin nka-t/\sqrt{|U|}\sin
(n-1)ka]/E_{k}$ with $E_{k}=\pm \sqrt{|U|+t^{2}/|U|-2t\cos ka}$.

\begin{figure}
\hspace*{-10mm}
\rotatebox{-90}{\includegraphics*[width=60mm]{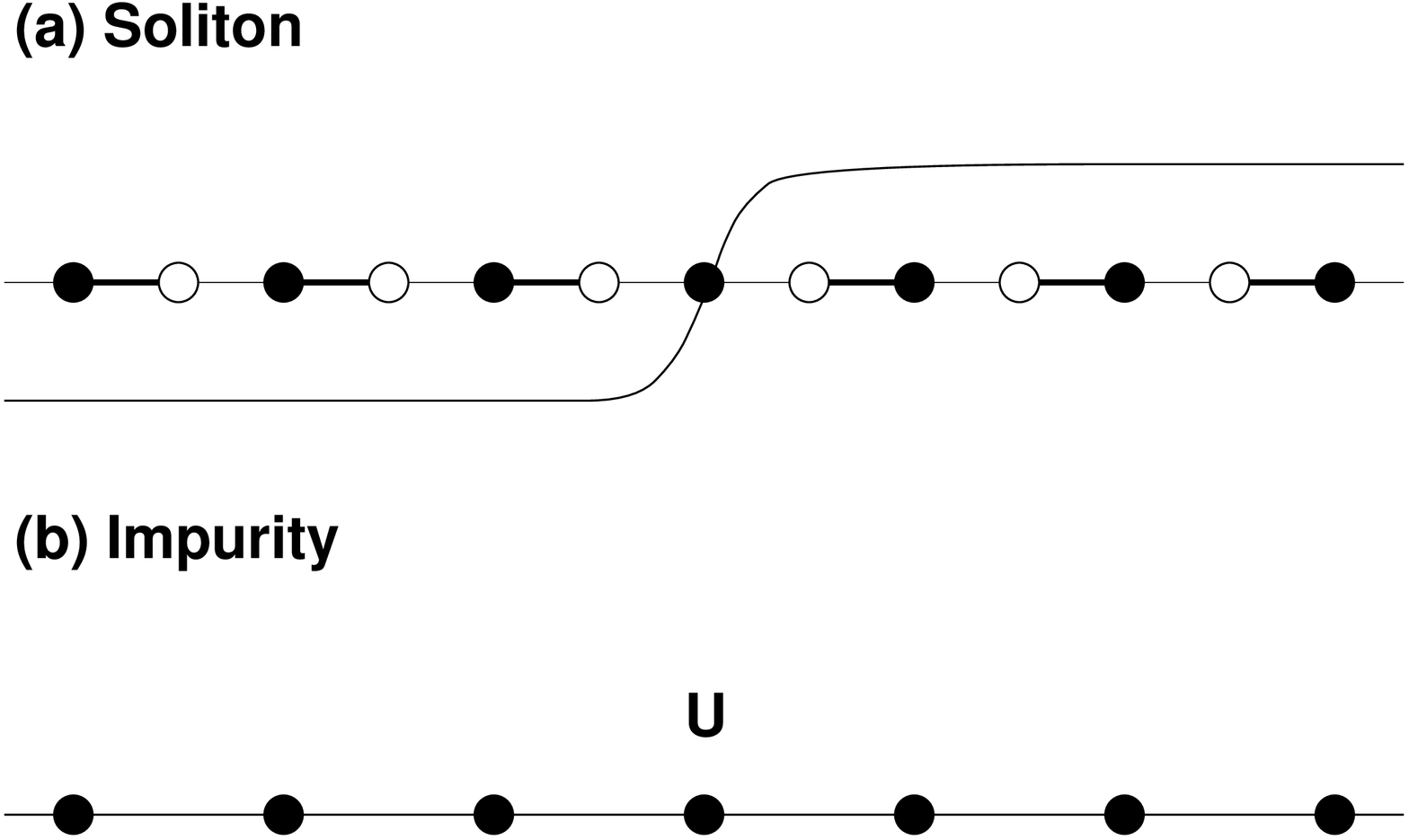}}
\caption{\small The equivalence of the electronic state in a SSH soliton
background (a) and  in an impurity background (b). The electronic
state for an impurity resides only on the black points.
In (a), the thin line represents hopping amplitude
 $t_1$, while the thick line represents hopping amplitude
 $t_2$. In (b), the hopping amplitude
is $t_1 t_2$. \label{SSHsol}
}
\end{figure}

One can also apply
the same approach to the bulk case.
In Fig.~\ref{SSHsol}(a), we show that
the soliton configuration
in polyacetylene discussed by Su, Schrieffer, and Heeger
(the SSH soliton)\cite{Su1,Su2}
is essentially a bulk impurity illustrated in Fig.~\ref{SSHsol}(b). The same
analysis leads to the results: For $U<0$,
$E_{imp} = -\sqrt{U^2+4t^2}$ and $\Psi_{imp} (n)
= \left[ |U|/(2t) - \sqrt{(U/2t)^2+1} \right]^{|n|}$,
while for $U>0$, $E_{imp} = \sqrt{U^2+4t^2}$, and $\Psi_{imp} (n)
= \left[ -|U|/(2t) +\sqrt{(U/2t)^2+1} \right]^{|n|}$. Here $n$ is measured
from the impurity site. Note that unlike the case for semi-infinite
chains, the impurity state for the bulk 1D chain exists for any $U$
and $t$.

The purpose of the above analysis is only for demonstration.
In principle, it can be also applied to cases when the potential
extends from the first to other lattice points in the Goodwin model.
For instance, one can include an additional term $V c_{2}^{\dagger}c_{2}$ in
the Goodwin model. In that case, the first and second bonds in
the corresponding $t_{1}$-$t_{2}$ model has to be different
and denoted as $t'_1$ and $t'_2$. One then obtains that
$t=t_1 t_2 = {t'}_1 {t'}_2$, $U={t'_1}^2 - {t_1}^2 -{t_2}^2$,
and $V={t'_2}^2 -{t_2}^2$.
Since for any changes of finite number of hoppings, the
energy of the edge mode stays at zero. The energy
of the impurity state is still given by $E_{imp}= -({t_1}^2+{t_2}^2)$.
Solving ${t_1}^2$ and ${t_2}^2$ in terms of $t$, $U$, and $V$
yields the energy of the impurity state.

We close this section by pointing out that the idea of mapping
impurity states to midgap states is quite general. In addition to
the simplest version of the Goodwin model discussed so far,
it also works for more general cases.
For instance, consider the generalized Goodwin model $\tilde{H}_G$
in which one introduces an impurity potential
at the 1st site of the $t_{1}$-$t_{2}$ model. This model can be
mapped to the effective Hamiltonian for the odd sites
of a $t'_{1}$-$t'_{2}$-$t'_{3}$-$t'_{4}$ model, i.e., the $p=4$
($H_4$) model. Specifically,
one finds that $\tilde{H}_G$ is equivalent to the block in ${H_4}^2$
that describes the odd sites; while its supersymmetric partner
is the original $t_{1}$-$t_{2}$ model with no impurity potential.
The impurity state of $\tilde{H}_G$ is identified as the midgap state of
the $H_4$ model at $E=0$. Simple calculation then yields
$E_{imp}={t_{2}}^2/(2U)+U/2+\sqrt{({t_{2}}^2/(2U)+U/2)^2+{t_{1}}^2}$.
Thus, one concludes that in general the impurity state of $H_p$ can be mapped to
the midgap state of $H_{2p}$.

\section{Summary and outlook}
\label{conclude}

In summary, in this work we have shown that the properties of
midgap states in semi-infinite
1D nanowires are dictated by a 
underlying discrete supersymmetry. 
This supersymmetric structure generalizes
the ordinary supersymmetric quantum mechanics 
and offers a new point of view toward the origin of edge states.
In the presence of hard-wall boundary condition ($\psi =0$), the
sublattice which directly connects to the hard-wall spans
the null space of the supersymmetric ground state.
As a consequence,
the energies of the midgap states
are determined by the eigenvalues of a reduced Hamiltonian, Eq.(\ref{mid}),
whose size is much smaller than
that of the original Hamiltonian.
This reduction in matrix size significantly reduces the computation cost for determining
the occurrence of the midgap
states and offers a way to manipulate them.
As an application, we investigate a structure with superlattice in hopping.
In this case, the number of edge states
is simply the period of the superlattice minus one.
Therefore, changing the period offers a way to control
the number of localized electronic states at the edge
of the nanowires.

While so far in this work we have not considered electron-electron
interactions, from adiabatic continuity, the results obtained here should still hold
for cases when the electron-electron
interactions are weak (so that quasi-particles are well defined).
Moreover, if the ``1D chain" results from
the reduction of a higher dimensional structure where interactions are
not important, then it is legitimate to ignore interactions in its effective 1D model.
This is of course not correct in truly 1D atomic chains
where it is known that interactions may dominate the physics and
the quasi-particle pictures
may fail. In this case, however, the states we obtain
can be used as the basis to express the full Hamiltonian (with
interactions) utilizing the relation $c_{i \sigma}^{^{\dagger }}=\psi
_{0}(i) c_{0 \sigma}^{^{\dagger }}+\sum_{E}\psi _{E}(i)
c_{E \sigma}^{^{\dagger }}$. When Coulomb interaction is included,
it reduces to the Anderson model\cite{Mahan} in which the edge state
acts as an impurity state. The scattering
of extended states by the edge states essentially causes the Kondo effect,
resulting in the zero-bias peaks near the Fermi energies
\cite{kondo}. On the other hand, the interaction also
correct the localized edge state. This is conventionally analyzed
in the Fano-Anderson model\cite{Mahan}, in which it is
known that as long as the new energy found remains
inside the gap, the corresponding state is a localized
state. In either of the above mentioned effects, our results will serve as
useful inputs for attaining the final corrections.

\begin{acknowledgments}
We gratefully acknowledge discussions with
Profs. Sungkit Yip, Hsiu-Hau Lin, and
T. K. Lee
and
support from the National Science Council of the Republic of China under
Grant Nos. NSC 92-2112-M-007-038.
\end{acknowledgments}

\appendix

\renewcommand{\theequation}{\Alph{section}$\;\arabic{equation}$}

\section{Superalgebra in infinite $p$-partite systems}
                                                                                
In this appendix, detailed superalgebra behind our generalized SUSY quatum
mechanics is presented. For an infinite $p$-partite system, after modulo $p$,
the system reduces to the set $\{ 1, 2, 3, \cdots,p \}$ with each number 
representing different sublattices. The reduced system is periodic
with $p+1$ being identified as $1$.
In this periodic space, 
a set of generators $\{Q_{nm};n,m=1,2,...,p \}$ 
can be defined. Here $Q_{nm}$ are $p\times p$ Hermitian matrices whose only
nonvanishing elements are $1$ in the $n$ th row and $m$ th column and the $m$
th row and $n$ th column. Note that $Q_{nn}$ has only one element, $1$,
in the $n$th element along the diagonal.
Obviously, when $n \neq m$, $Q_{nm}$ permutes the subwavefuncions $
\phi_{n}$ and $\phi _{m}$; when combined with the hopping strength
$t_{nm}$, in addition to permutation, it also rescales the wavefunctions. 
The Lie algebra formed by $Q_{nm}$ is a superalgebra
because the anticommutator is necessary in
order to be closed. The followings are nontrivial
commutation relations: $\{Q_{lm},Q_{mn}\}=Q_{ln}$ for $l\neq n$ , $
\{Q_{nm},Q_{nm}\}=2Q_{nn}+2Q_{mm}$ for $n \neq m$, 
and $[Q_{nn},Q_{mm}]=\delta _{nm}Q_{n}$; 
all the other commutators are zero. It is
straightforward to check that for infinite systems, the SUSY Hamiltonian $
H_{S}$ defined in Sec.III commutes with all $Q_{nm}$ 
even if the operation of modulo $p$ is not performed.
                                                                                
\section{Derivation of $[H_{S},Q_{p}]=0$}

In this appendix, we outline the proof of $[H_{S},Q_{p}]=0$ for
semi-infinite chains. We first write $Q_{p}=q_{p}+q_{p}^{^{\dagger }}$ where
$q_{p}$ is obtained by setting the last column in $Q_{p}$ to zero in
Eq.(\ref{super1}). It is then suffice to prove $[H_{S},q_{p}]=0$. We note that $H_{S}$
has the following generic form
\begin{equation}
H_{S}=\left(
\begin{array}{ccccc}
\mathbf{S}_{11} & \mathbf{S}_{12} & \mathbf{S}_{13} & \cdot \cdot  & \mathbf{
0} \\
\mathbf{S}_{21} & \mathbf{S}_{22} & \mathbf{S}_{23} & \cdot \cdot  & \mathbf{
0} \\
\mathbf{S}_{31} & \mathbf{S}_{32} & \mathbf{S}_{33} & \cdot \cdot  & \cdot
\\
\cdot \cdot  & \cdot \cdot  & \cdot \cdot  & \mathbf{S}_{p-1,p-1} & \mathbf{0
} \\
\mathbf{0} & \mathbf{0} & \cdot \cdot  & \mathbf{0} & \mathbf{0}
\end{array}
\right) .+H_{S}^{0}\equiv \left(
\begin{tabular}{ccc|c}
&  &  & $\mathbf{0}$ \\
& $\mathbf{H}^{+}$ &  & $\cdot$  \\
&  &  & $\mathbf{0}$ \\ \hline
$\mathbf{0}$ & $\cdot$  & $\mathbf{0}$ & $\mathbf{H}^{-}$
\end{tabular}
\right) \label{block1}
\end{equation}
Here $H_{S}^{0}$ has the same form as that of the SUSY Hamiltonian 
for the corresponding infinite chain except that semi-infinite lattice points
are removed, hence it is block-diagonal with the form:
 $(H_{S}^{0})_{nm}=\mathbf{S}_{0}\delta _{nm}$, where
$n$ and $m$ are the block indices and
\begin{equation}
\mathbf{S}_{0}=\left(
\begin{array}{ccccc}
0 & t & 0 & \cdot \cdot  & 0 \\
t & 0 & t & \cdot \cdot  & 0 \\
0 & t & 0 & \cdot \cdot  & \cdot  \\
\cdot \cdot  & \cdot \cdot  & \cdot \cdot  & \cdot \cdot  & t \\
0 & 0 & \cdot \cdot  & t & 0
\end{array}
\right) \label{block2}
\end{equation}
and $t=t_{1}t_{2}t_{3}\cdot \cdot t_{p}$. 
The block matrix $\mathbf{S}_{nm}$ represents $\it{missing}$
hopping amplitudes that goes between sublattices $m$ and $n$
due to the presence of the boundary point at $i=0$.
When computing $[H_{S},q_{p}]$, one needs to compute
$[\mathbf{S}_{0},{\bf A}_{1p}^{^{^{\dagger }}}]$,
${\bf A}_{1p}^{^{^{\dagger }}} \cdot \mathbf{S}_{1m}$ 
and ${\bf A}_{p-1,p}^{^{^{\dagger }}} \cdot \mathbf{S}_{p-1,m}$,
thus only $\mathbf{S}_{1m}$ and $\mathbf{S}_{p-1,m}$ are needed.
It is straightforward to show that 
$[\mathbf{S}_{0},{\bf A}_{1p}^{^{^{\dagger }}}]$ has only one
element $-t t_{p}$, which is
the 1st element along the diagonal ($\equiv \delta_{11}$).
To obtain $\mathbf{S}_{nm}$, one needs to multiply $H_{p}$ to
itself by $n$ ($\leq p$) times because $H_{S}$ is a polynomial of
$H_{p}$ containing at most $p$th power of $H_{p}$. 
Now the multiplication of $H_{p}$ to itself
$n$ times effectively hops a particle $n$ times. Since $n \leq p$,
only when the particle starts from the lattice points 
$1 \leq i \leq p-1$, it will have chance to visit $i=0$ and
thus will have missing paths when $i=0$ is removed. It implies
that all $\mathbf{S}_{nm}$ have only one element, which is also
the 1st element along the diagonal.
In addition, $\mathbf{S}_{p-1,m}=0$ for $m \geq 2$.
As a result, by using Eq.(\ref{Ap}), we obtain
${\bf A}_{1p}^{^{^{\dagger }}} \cdot \mathbf{S}_{1m} =0$ and hence
we only need to compute $\mathbf{S}_{p-1,1}$.
Since the only missing path between the point $1$ and
the point $p-1$ is $1 \rightarrow 0 \rightarrow 1
\rightarrow 2 \cdots \rightarrow p-1$, we find that
$\mathbf{S}_{p-1,1}$ is
$-{t_p}^2 t_1 t_2 \cdots t_{p-1} \delta_{11}$. 
This result, when combined with 
$[\mathbf{S}_{0},{\bf A}_{1p}^{^{^{\dagger }}}]=- t t_p \delta_{11}$,
we finally obtain $[H_{S},Q_{p}]=0$.


\begin{references}
\bibitem{para1} The construction of $H_S$ essentially used the Hamilton-Cayley
theorem often quoted in elementary linear algebra. Here for
the purpose of reference,
we list some of its forms:
$p=3$: $H_S={H_p}^3 - ({t_1}^2+{t_2}^2+{t_3}^2) H_p$; $p=4$:
$H_S={H_p}^4 - ({t_1}^2+{t_2}^2+{t_3}^2 +{t_4}^2) {H_p}^2 +{t_2}^2 {t_4}^2
+{t_1}^2 {t_3}^2$.
Clearly, for higher $p$, it involves higher powers of $H_p$.
\bibitem{complete} Because $\phi_p$ is an eigenstate to $\mathbf{H}^{-}$ 
with eigenvalue $E_S$ and $\mathbf{H}^{-}$ is a uniform hopping model,
$\phi_p$ is either in the form of $sin(npka)$ or simply zero. 
Since $\phi_p = sin(npka)$ is an extended state, the only
possibility for accommodating the midgap states is $\phi_p=0$. Therefore,
$\phi_p=0$ and the following construction exhaust all possibilities.
\bibitem{index} This is equivalent to find the number of midgap states.
Using $\mathbf{H}^{+}$ and $\mathbf{H}^{-}$,
the number of midgap states can be formally counted.
For instance, if one defines
the index as
$\mathrm{ind}(z)\equiv \mathrm{Tr}\left( \frac{z}{z+\mathbf{H}^{+}}-\frac{z}{%
z+\mathbf{H}^{-}}\right) +\mathrm{Tr}\left( \frac{\mathbf{H}^{+}}{z+\mathbf{H%
}^{+}}-\frac{\mathbf{H}^{-}}{z+\mathbf{H}^{-}}\right) $, $\mathrm{ind}(0)$
gives the number of midgap states.
\end{references}
\end{document}